\documentstyle[aps,twocolumn,graphicx,amsfonts,amsmath,array,cite,citesort,float]{revtex}
\restylefloat{figure}
\newcommand{\lz}{\ell=0}
\newcommand{\li}{\ell=\infty}
\newcommand{\ldep}{(\ell)}
\newcommand{\crit}{\text{c}}
\newcommand{\diff}{\text{d}}
\newcommand{\eff}{\text{eff}}

\newcommand{\hc}{\text{h.c.}}
\newcommand{\sw}{\text{sw}}

\newcommand{\gc}{g_{\crit}}

\newcommand{\om}{\omega}
\newcommand{\omz}{\om_0}
\newcommand{\ome}{\om_{\eff}}
\newcommand{\oml}{\om\ldep}
\newcommand{\mue}{\mu_{\eff}}
\newcommand{\mul}{\mu\ldep}

\newcommand{\dom}{/\om}
\newcommand{\domz}{\dom_0}

\newcommand{\al}{\alpha}
\newcommand{\ale}{\al_{\eff}}

\newcommand{\alc}{\al_{\crit}}

\newcommand{\la}{\lambda}

\newcommand{\lap}{\la_{\text{ph}}}
\newcommand{\lapz}{\la_{\text{ph,0}}}

\newcommand{\etal}{\eta\ldep}

\newcommand{\bd}{b^{\dagger}}
\newcommand{\bi}{b_i}
\newcommand{\bdi}{b_i^{\dagger}}

\newcommand{\h}{{\mathcal{H}}}
\newcommand{\hl}{\h\ldep}
\newcommand{\hs}{\h_{\text{S}}}
\newcommand{\hb}{\h_{\text{B}}}
\newcommand{\hbw}{\h_{\text{B}}^{(\om)}}
\newcommand{\hbm}{\h_{\text{B}}^{(\mu)}}
\newcommand{\hsb}{\h_{\text{SB}}}
\newcommand{\hd}{\h_{\diff}}
\newcommand{\he}{\h_{\text{eff}}}
\newcommand{\hsl}{\h_{\text{S}}\ldep}
\newcommand{\hbl}{\h_{\text{B}}\ldep}
\newcommand{\hsbl}{\h_{\text{SB}}\ldep}
\newcommand{\hdl}{\h_{\diff}\ldep}
\newcommand{\qz}{{\mathcal{Q}}}

\newcommand{\opl}{{\mathfrak{L}}}

\newcommand{\opa}{{\mathcal{A}}}
\newcommand{\opad}{\opa^{\dagger}}
\newcommand{\opadi}{\opad_i}
\newcommand{\opai}{\opa_i}

\newcommand{\opail}{\opai\ldep}
\newcommand{\opadil}{\opadi\ldep}

\newcommand{\opspure}{{\mathbf{S}}}
\newcommand{\opsgen}[1]{\opspure_{#1}}
\newcommand{\ops}[1]{\opsgen{i#1}}
\newcommand{\sprgen}[2]{\opsgen{#1}\opsgen{#2}}

\newcommand{\spr}[2]{\sprgen{i#1}{i#2}}

\newcommand{\kpr}[3]{\ops{#1}(\ops{#2}\times\ops{#3})}

\newcommand{\vprgen}[4]{(\opsgen{#1}\times\opsgen{#2})(\opsgen{#3}\times\opsgen{#4})}
\newcommand{\dell}{\diff\ell}
\newcommand{\diffell}[1]{\frac{\diff#1}{\dell}}

\newcommand{\limellinf}{\ell\rightarrow\infty}
\newcommand{\onehalf}{1/2}
\newcommand{\oh}{\onehalf}

\newcommand{\spinonehalf}{S\!=\!\oh}

\tighten
\begin{document}
\setlength{\unitlength}{1mm}


\title{Effective Spin Models for Spin-Phonon Chains by Flow Equations}
\author{Carsten Raas\thanks{e-mail: cr@thp.uni-koeln.de},
Alexander B\"uhler\thanks{e-mail: ab@thp.uni-koeln.de},
and G\"otz S. Uhrig\thanks{e-mail: gu@thp.uni-koeln.de}}

\address{Institut f\"ur Theoretische Physik, Universit\"at zu
  K\"oln, Z\"ulpicher Str.~77, D-50937 K\"oln, Germany\\[1mm]
  {\rm(\today)} }

\maketitle
\begin{abstract}
We investigate the anti-adiabatic limit of an anti-ferromagnetic $\spinonehalf$
Heisenberg chain coupled to Einstein phonons. The flow equation method is used
to decouple the spin and the phonon part of the Hamiltonian. 
In the effective spin model long range spin-spin interactions are generated.
We determine the phase transition from a gapless state to a gapped (dimerized)
phase, which occurs at a non-zero value of the spin-phonon coupling.
In the effective phonon sector a phonon hardening is observed.
\end{abstract}
\narrowtext
\noindent\emph{Dedicated to Professor E. M\"uller-Hartmann on the occasion of
  his 60th birthday.}
\section{Introduction}
\label{capInt}
The interest in the spin-Peierls (SP) transition \cite{bray83} has been
renewed by the discovery of the first inorganic SP substance CuGeO$_3$ 
\cite{hase93a,bouch96}. The SP instability of quasi one-dimensional spin
systems is the analogon of the Peierls instability of quasi one-dimensional
metals towards lattice modulations with a wave vector $2k_F$. Due to the
coupling of the lattice degrees of freedom to the $d=1$ magnetic degrees of
freedom the system can lower its energy by undergoing a phase transition into
a dimerized state. The loss in elastic energy is overcompensated by the gain
in the magnetic energy.

Most theoretical studies of SP transitions rely on an \emph{adiabatic}
treatment of the phonons (e.\,g.\@ Ref.\@ \cite{cross79}). This means that the
energy of the phonons, that are associated with the lattice distortion, should
be low compared with the spin-spin exchange coupling. In addition to the
occurrence of a \emph{pre-existing soft phonon}, one observes a phonon
softening in the organic SP compounds. Furthermore, the phonon energy $\omega$
should be small compared with the resulting gap $\Delta$ in the spin
excitation spectrum \cite{caron96}.

It has been recently emphasized \cite{uhrig98b}, that CuGeO$_3$ does not
fulfill these conditions. Even more, a \emph{phonon hardening} by about
$5\%-6\%$ is observed \cite{brade98a}. Uhrig \cite{uhrig98b} has developed a
flow equation approach \cite{wegne94} which is not founded on adiabaticity.
The comparison with the DMRG data of Bursill et al.\@ \cite{bursi99} has shown
that the flow equation method works well, especially in the limit of small
$J\dom$ (cf.\@ Fig.~1 in Ref.\@ \cite{bursi99}). In this paper we extend
this approach. Details will be given below.

\section{Flow equation approach}
\label{capFlow}
The method of our choice is the flow equation approach introduced by Wegner
\cite{wegne94}. This method is similar to Fr\"ohlich's approach
\cite{frohl52}, which projects a Hamilton operator to an effective Hamiltonian
in one step applying an unitary transformation. Wegner modified the Fr\"ohlich
transformation as he uses an infinite number of infinitesimal unitary
transformations.

The model we study reads
\begin{mathletters}
\begin{eqnarray}
  \hl & = & \hs + \hsb + \hb \label{eqHl}\\
  \hs & = & \sum_{i} (J_1 \ldep\spr{}{+1} + J_2\ldep\spr{}{+2}) \label{eqHs}\\
  \hb & = & \sum_i\big[\oml\bd_i\bi+\frac{\mul}{2}(\bdi\bdi+\bi\bi)\big]
          \label{eqHb}\\
  \hsb & = & \sum_i (\opail\bdi+\opadil\bi)\ . \label{eqHsb}
\end{eqnarray}
\end{mathletters}
Here $\ops{}$ stands for the $\spinonehalf$ spin operator on site $i$ and
$\bi$ ($\bdi$) destroys (creates) a phonon on site $i$. $\hs$ is the
Hamiltonian of a frustrated spin chain or $J_1$-$J_2$ model. If the
frustration parameter $\al\equiv J_2/J_1$ exceeds a critical value of
$\al=\alc=0.241167(5)$, then the model undergoes a quantum phase transition
from a gapless state to a gapped phase \cite{affle89,okamo92,egger96}.
We start with a Hamilton operator (\ref{eqHl}) consisting of
pure spin ($\hs$) and phonon ($\hb$) parts and a spin-phonon coupling term
$\hsb$ and use flow equations to decouple the spin-phonon system.

The continuous transformation is parameterized by a flow parameter
$\ell\in[0,\infty]$. So $\h(\lz)$ stands for the original Hamiltonian as
given. We will end up with the effective Hamiltonian $\h(\li)$ which is more
``diagonal'' in the way that the direct spin-phonon coupling has been rotated
away. The infinitesimal generator $\etal$ defines the unitary transformation
via
\begin{eqnarray}
  \diffell{\hl}=[\etal,\hl]\ .
  \label{eqFlow}
\end{eqnarray}
A choice of $\etal$ proposed by Wegner is 
\begin{eqnarray}
  \etal=[\hdl,\hl]\ ,
  \label{eqEtaGen}
\end{eqnarray}
where $\hd$ is the part of the Hamiltonian which is regarded as diagonal. By
choosing $\hdl\equiv\hsl+\hbl$ and the ``off-diagonal'' interaction part as
$\h_{\text{od}}\ldep\equiv\hsbl$, our generator reads
$\etal\equiv[\hd,\h_{\text{od}}] = [\hs+\hb,\hsb]$. In Chapter \ref{capMod} we
present an alternative choice of $\eta$, while the canonical version
(\ref{eqEtaGen}) is used throughout the main part of this paper.

As initial conditions we choose
\begin{eqnarray}
  J_1(0)=J \quad \text{and} \quad J_2(0)=0.
\end{eqnarray}
So the untransformed pure spin part is the normal Heisenberg Hamiltonian. As
we will see later on, for non-zero values of the flow parameter $\ell$ long
range spin-spin interactions arise. As a consequence, the next-nearest
neighbour coupling constant $J_2$ in the effective spin model is finite. The
phonon part $\hb=\hbw+\hbm$ includes a non-boson-number conserving term
$\hbm$. This choice is due to the fact, that the generator
$\eta=[\hs+\hbw,\hsb]$ would produce terms proportional to $\bi\bi$
($\bdi\bdi$) anyway. So for our purpose it is convenient to embed this terms in
$\h$ and renormalize $\mul$. Henceforth, we choose
\begin{eqnarray}
  \om(0)\equiv\omz=1 \quad \text{and} \quad \mu(0)=0
\end{eqnarray}
and give all coupling constants in units of $\omz$.

$\hsb$ describes the coupling between spin and phonon system. The spin-phonon
coupling mainly influences neighbouring sites. Hence the coupling operator
$\opai(0)$ typically consists of nearest neighbour spin products. An
appropriate choice for $\opai(0)$ is the difference coupling
\begin{eqnarray}
  \opai(0)=g(\spr{}{+1}-\spr{}{-1})\ ,
  \label{eqOpA}
\end{eqnarray}
which does not contribute to the undistorted phase due to the translational
invariance \cite{uhrig98b}, i.\,e.\@ $\langle\opai(0)\rangle$ vanishes.
According to Ref.\@ \onlinecite{wegne94}, $\opai$ should be normal-ordered,
i.\,e.\@ $\opai(0)\rightarrow\opai(0)-\langle\opai(0)\rangle$. This is
guaranteed by the specific choice (\ref{eqOpA}). By choosing different
coupling operators, various mechanisms of how the lattice distortions
influence the exchange integral $J$ can be investigated. In the case of the
difference coupling, the exchange coupling directly depends on the
``position'' of a spin between the neighbouring ones. For CuGeO$_3$ this
corresponds to changes of two neighbouring Cu--O--Cu binding angles,
i.\,e.\@ one of them is enlarged on expense of the other one.

Sometimes a local coupling $\hsb^{\text{loc}}=g\sum_i\spr{}{+1}(\bdi+\bi)$ is
used in connexion with CuGeO$_3$ (see e.\,g.\@ Ref.\@
\onlinecite{kuehne99,weisse99b,trebst99cm}): Single harmonic degrees of freedom
directly modify the magnetic interaction. In Ref.\@ \onlinecite{trebst99cm} the
gap due to dimerization was investigated by means of a high order perturbation
expansion. A flow equation approach to this coupling type will be presented
elsewhere \cite{raas01b}.

Applying the unitary transformation, not only the coupling constant in
(\ref{eqOpA}) changes ($g\to a\ldep$). Additionally longer range
and multiple site interactions arise. Their $\ell$-dependent coupling
constants are denoted by $b_n\ldep$. The aim of this approach is to
disentangle phonons and spins. Hence in the limit $\limellinf$ the coupling
operator $\opail$ vanishes, i.\,e.\@ $a(\infty)=0$ and $b_n(\infty)=0$.

We introduce the Liouville operator $\opl$ for the commutation with $\hs$:
$\opl\mathcal{O}:= [\hs, \mathcal{O}]$. In extension to
Ref.\@ \onlinecite{uhrig98b} the Liouville operator $\opl\ldep$, the phonon
frequencies $\oml$ and $\mul$ are $\ell$-dependent. We choose for the
generator $\eta$
\begin{eqnarray}
\etal = \sum_i\Big\{
    &\big[(\opl+\om)\opai -\mu\opadi\big]&\bdi+\nonumber\\
    &\big[(\opl-\om)\opadi+\mu\opai \big]&\bi
    \Big\}
\end{eqnarray}
according to $\eta=[\hs+\hb,\hsb]$. The general flow equation (\ref{eqFlow})
leads to a flow equation for the coupling operator $\opail$ by investigating
the terms linear in phonon operators of $[\etal,\hl]$
\begin{equation}
  \label{eqDGLA}
  \diffell{\opai} = -
  \Big\{\big[(\opl+\om)^2 - \mu^2\big]\opai - 2\mu\opl\opadi\Big\}\ .
\end{equation}
Due to the complicated structure of Eq.~(\ref{eqDGLA}) we restrict ourselves
to a subspace of operators, which are allowed to appear in $\opail$. A
systematic choice is to consider all terms arising in $\opai(0)$ and
$\opl\opai(0)$. The following table summarizes the resulting set of operators.

\begin{center}
\begin{tabular}{|c|c|} \hline\label{tabOperatorsM5}
    coefficient & spin operator \\ \hline
    $a\ldep$   & $\spr{}{+1} - \spr{-1}{}$\\
    $b_0\ldep$ & $i \kpr{-1}{}{+1}$\\
    $b_1\ldep$ & $i [\kpr{}{+1}{+2} + \kpr{-2}{-1}{}]$\\
    $b_2\ldep$ & $i [\kpr{}{+1}{+3} + \kpr{-3}{-1}{}]$\\
    $b_3\ldep$ & $i [\kpr{+2}{}{-1} + \kpr{-2}{}{+1}]$\\
    \hline
\end{tabular}
\end{center}

\noindent
In next order in $\opl$ the operators arising in $\opl^2\opai(0)$ are
included leading to 14 additional three spin operators of the type
$\vprgen{i}{j}{k}{l}$ and five additional long range two spin operators.
Eq.~(\ref{eqDGLA}) can now be divided in a set of differential equations
for the spin-phonon coupling constants $a\ldep$ and $b_n\ldep$.

The truncation of the coupling operator $\opai$ in order $n$ (by constructing
the operator space with all terms appearing in $\opai(0)$, $\opl\opai(0)$,
$\opl^2\opai(0)$, $\ldots$, $\opl^n\opai(0)$) neglects effects of order $n+1$
in $J$. We note here, that in this step no expansion in $g$ occurs. As the
contribution to the flow of the neglected operators corresponding to higher
order contributions in $J$ is small compared to the one of the leading order
terms, we expect the flow equations to converge for $\limellinf$. This
actually is the case except for large values of $J\dom$.

For the phonon frequencies we find
\begin{mathletters}
\begin{eqnarray}
  \label{eqDGLom}
  \diffell{\om} &=& \Big\langle\sum_i\big(
    [(\opl+\om)\opai,\opadi] +
    [\opai,(-\opl+\om)\opadi]\big)
    \Big\rangle\\
  \label{eqDGLmu}
  \diffell{\mu} &=& 2 \Big\langle\sum_i
  [\opl\opai-\mu\opadi,\opai]
  \Big\rangle\ .
\end{eqnarray}
\end{mathletters}
Only nearest and next-nearest neighbour spin products are kept and their mean
field value, denoted by the outer brackets $\langle\cdot\rangle$, is inserted.

The flow equations for $J_1\ldep$ and $J_2\ldep$ can be extracted
from $[\eta,\hsb]$
\begin{eqnarray}
  \label{eqDGLJ12}
  &\sum_i&\Big\{
  \big[(\opl+\om)\opai,\opadi\big] + \hc
  \Big\} \left\langle\bd b\right\rangle\nonumber\\
  + &\sum_i&\Big\{
  \big[\opl\opai-\mu\opadi,\opai\big] + \hc
  \Big\}\left\langle\bd\bd\right\rangle\\
  - &\sum_i&\Big\{
  \opadi\big[(\opl+\om)\opai-\mu\opadi\big] + \hc    
  \Big\}\nonumber\ .
\end{eqnarray}
The quadratic boson terms are replaced by their expectation values. This is
systematic in the sense of an expansion in $g$. This mean-field step neglects
fluctuation effects of order $g^2$. As these terms only arise in $g^2$-terms,
the total error is of order $g^4$ (cf.\@ Ref.\@ \cite{uhrig98b}). We note here,
that these expectation values are $\ell$-dependent due to the
$\ell$-dependence of $\om$ and $\mu$. Finally, $\diffell{J_1}$
($\diffell{J_2}$) is equal to the (next) nearest neighbour part of
(\ref{eqDGLJ12}), respectively.

\begin{figure}[ht]
\begin{center}
  \includegraphics[width=80mm]{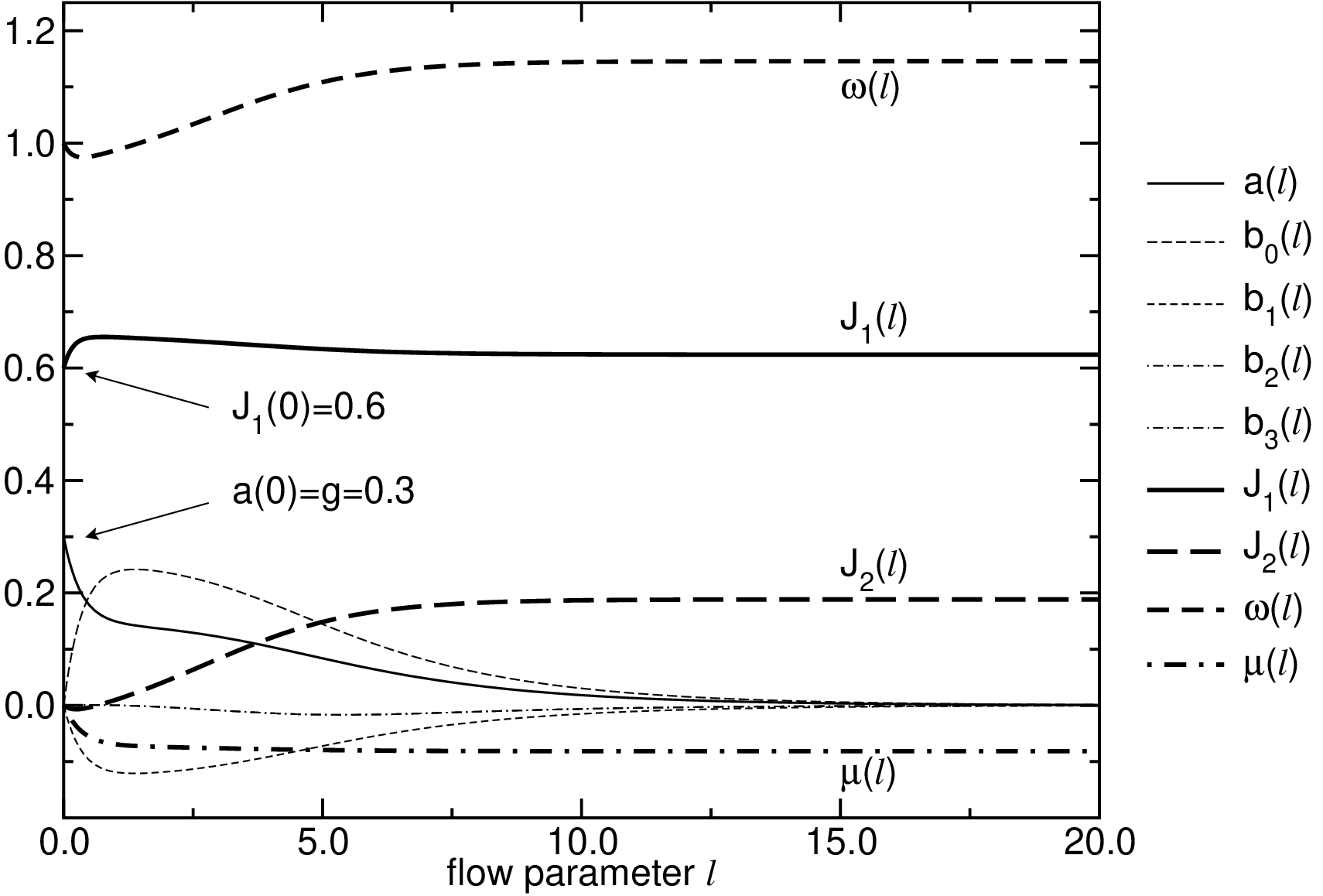}
\end{center}
\caption{\footnotesize{$\ell$-dependent flow of the the coupling constants,
    the phonon frequencies and the coefficients in the ansatz for the coupling
    operator $\opa$ in units of $\omz$.}
  \label{figFlow}}
\end{figure}

\section{Results}
\label{capRes}
Having collected all flow equations, the set of differential equations (9 in
order $\opl$ and 28 in order $\opl^2$) have to be solved numerically. This is
done via a standard fifth-order Runge-Kutta routine with adaptive step size
control. The change of the $\ell$-dependent coupling constants for two
succeeding Runge-Kutta steps is computed for all couplings. Convergence is
assumed if the sum of the changes squared is less than $10^{-9}$ in units of
$\omz$. Fig.~\ref{figFlow} shows for $J\domz=0.6$ and $g\domz=0.3$ the flow
of all 9 parameters. For intermediate values of $\ell$ the additional four
spin coupling terms arise ($b_n\ldep\neq0$). As $\ell$ becomes larger, all
spin-phonon coupling constants vanish, i.\,e.\@ $\hsb\to0$. $J_1$ and $J_2$
reach their effective values in the limit $\limellinf$ and thus give rise to a
frustration $\ale=J_2^{\eff}/J_1^{\eff}$ in the decoupled spin model. This
procedure is very intuitive in the sense that a flow diagram like
Fig.~\ref{figFlow} shows, how the dressing of the spins with phonons induces a
frustration $\ale>0$.

\begin{figure}[t]
\begin{center}
  \includegraphics[width=80mm]{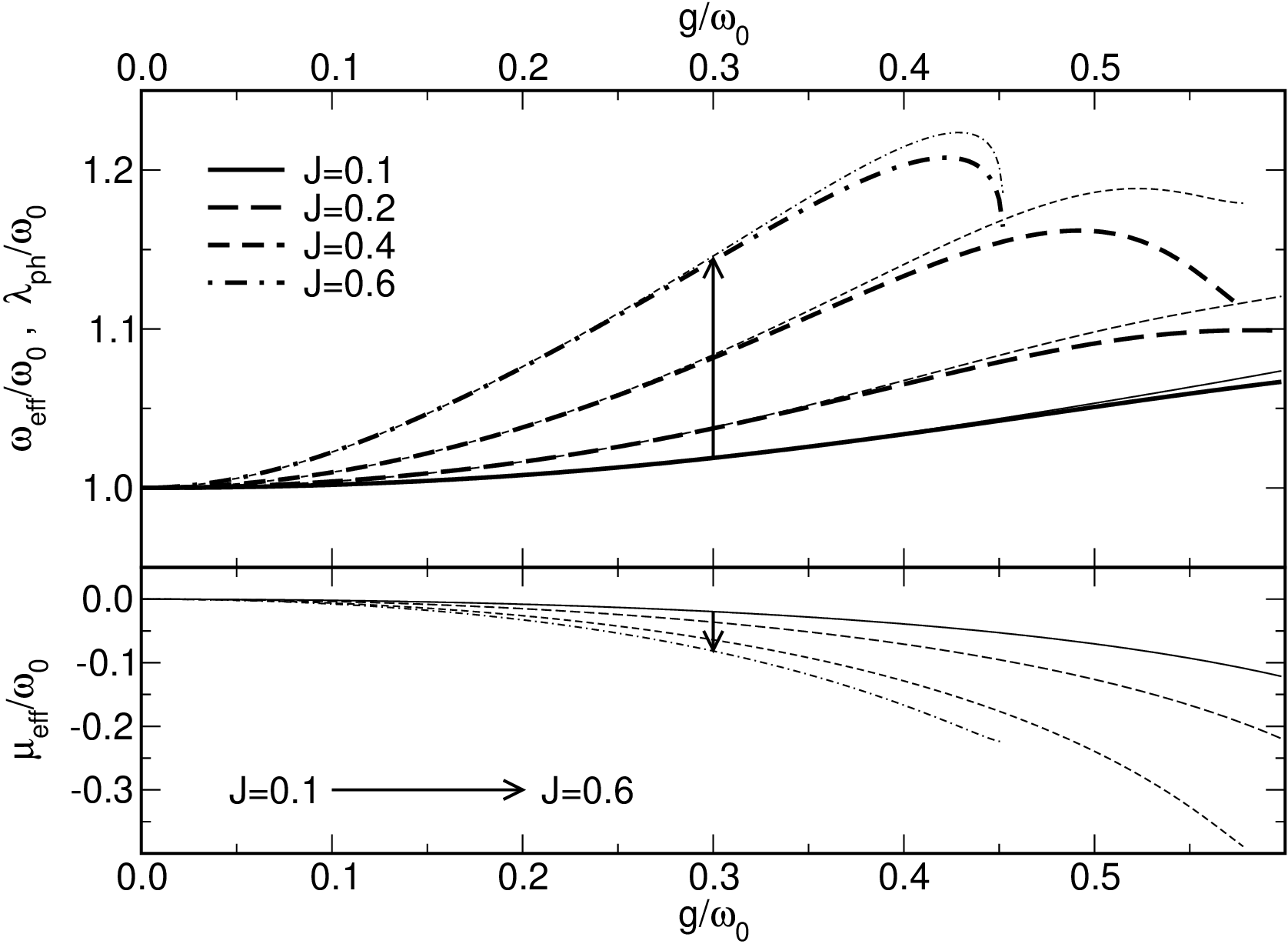}
\end{center}
\caption{\footnotesize{Effective phonon frequencies $\ome$ (thin lines in the
    upper part), $\mue$ (lower part) and $\lap$ (thicker lines) for fixed
    values of the initial next-nearest neighbour coupling $J=J_1(\lz)$.
    $\lap\equiv\sqrt{\ome^2-\mue^2}$ is the phonon frequency in the
    diagonalized version of the pure phonon part $\hb$.}
  \label{figomvsg}}
\end{figure}

As a by-product, we find an increased phonon frequency in the effective phonon
part of the model. This phonon hardening can be interpreted qualitatively as a
level repulsion between the high energy phonon system and the low energy spin
system. Recently, it has been pointed out by Gros and Werner \cite{gros98}
that this feature is consistent with the RPA-approach by Cross and Fisher to
the spin-Peierls transition. Fig.~\ref{figomvsg} shows for four values of $J$
the effective phonon frequencies in dependence of $g$. As an appropriate
measure of the phonon hardening, the frequency
$\lap\equiv\sqrt{\ome^2-\mue^2}$ of the diagonalized version
$\hb\to\sum_i\lap\beta^\dagger_i\beta_i+\lapz$ is depicted.

\begin{figure}[t]
\begin{center}
  \includegraphics[width=80mm]{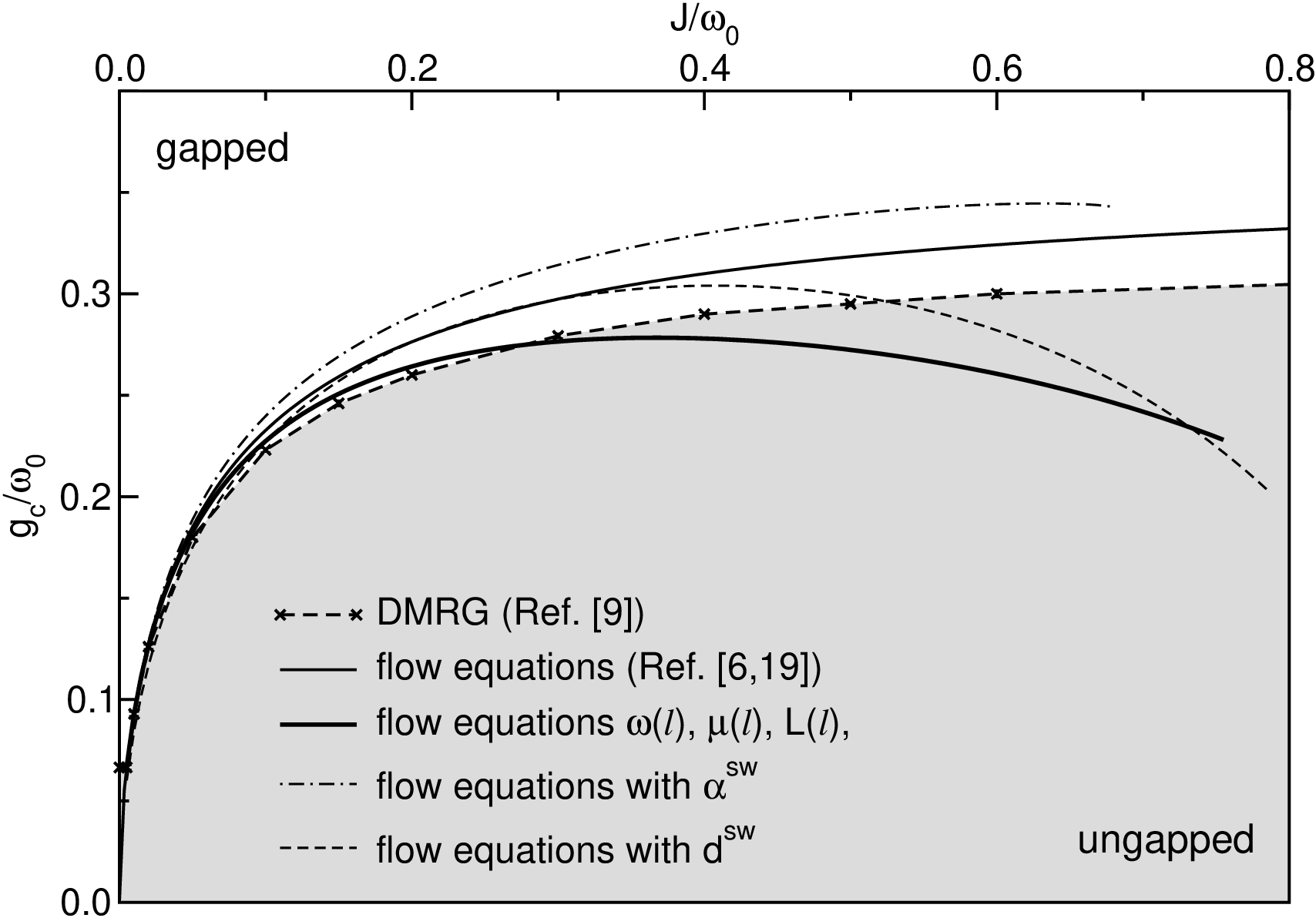}
\end{center}
\caption{\footnotesize{Zero temperature phase diagram of the spin-Peierls
    anti-ferromagnetic chain of spins interacting with phonons. For small
    values of the spin-phonon coupling $g\domz$ the system is gapless. For
    large $g\domz$ the system is dimerized and has an energy gap. The curves
    denoted with $\al^{\sw}$ and $\diff^{\sw}$ are the results of two modified
    approaches and will be discussed in Chapter \ref{capMod}.}
  \label{figgcvsjour}}
\end{figure}

\begin{figure}[t]
\begin{center}
  \includegraphics[width=80mm]{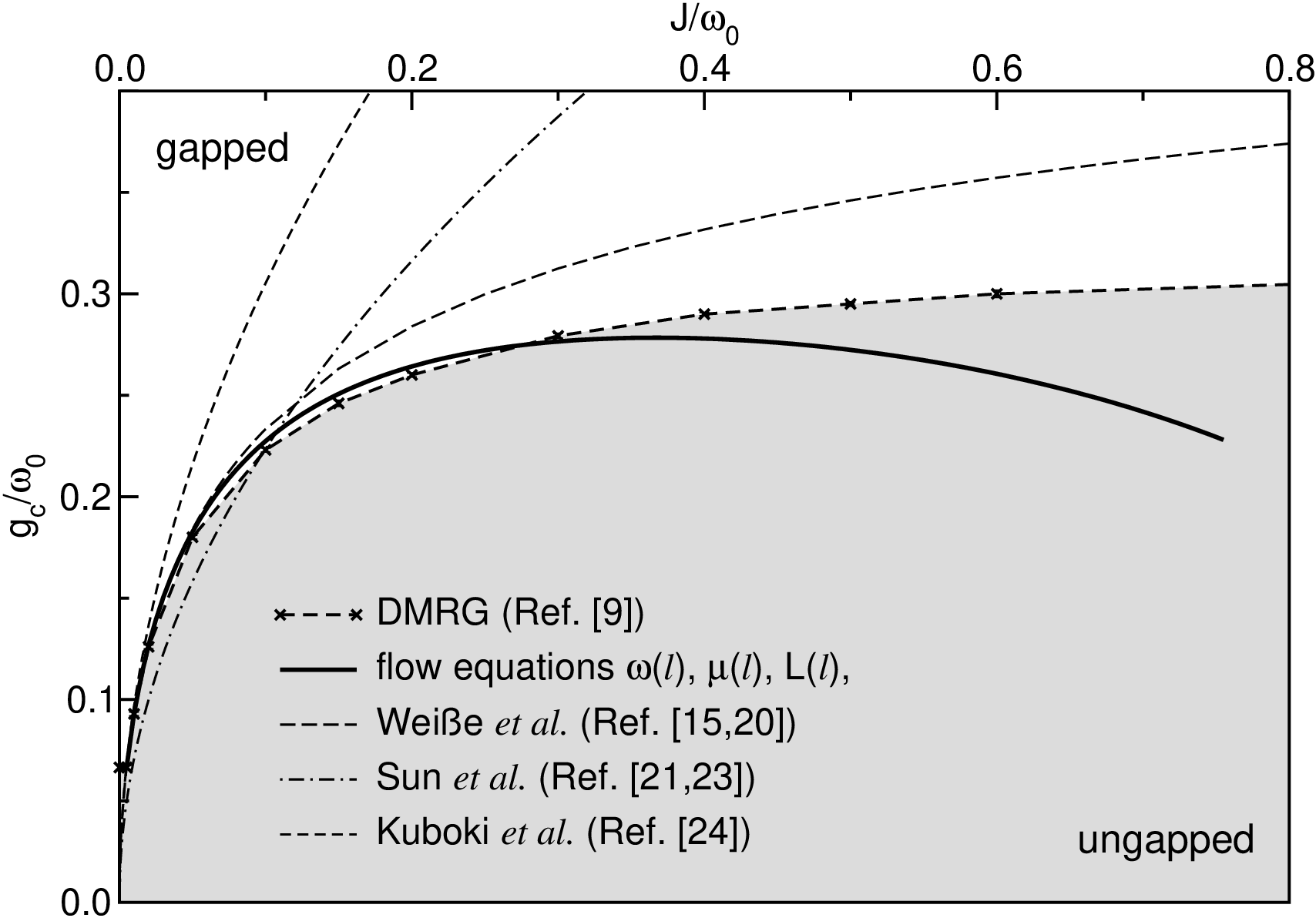}
\end{center}
\caption{\footnotesize{Phase diagram like Fig.~\ref{figgcvsjour}. Various
    previous results are compared with the phase transition lines obtained via
    flow equations or DMRG.}
  \label{figgcvsjother}}
\end{figure}

The phase diagram of the spin-Peierls anti-ferromag\-netic chain of spins
interacting with phonons has been investigated by Bursill et al.\@ via DMRG
\cite{bursi99}, by Uhrig via flow equations in leading order in $g\dom$ and
$J\dom$ \cite{uhrig98bremark}, by Wei\ss{}e et al.\@ via fourth order
perturbation theory and Lang-Firsov transformation \cite{weisse99bremark} and
by Sun et al.\@ via mean field and renormalization group methods \cite{sun00}.
In Fig.~\ref{figgcvsjour} results of various flow equation approaches are
shown. In Fig.~\ref{figgcvsjother}, we compare our results to those of the
other approaches.

Our calculation is based on the idea that the transition into the ordered
phase does not occur due to the softening of a phonon but due to the tendency
of the effective spin model towards dimerization. Phonon-induced frustration
above its critical value $\ale>\alc$ drives the dimerization. The phase
transition line is therefore determined by solving
\begin{eqnarray}
  \alc-\big(\lim_{\limellinf}\al\big)\big|_{g,J} = 0
\end{eqnarray}
for fixed values of $J$ with respect to $g$. In this way we find the critical
spin-phonon coupling $\gc$ in dependence of the nearest neighbour spin
coupling $J$. For values of $J\dom$ up to $0.35$ the extended flow equation
approach improves the results from Ref.\@ \cite{uhrig98b}. For intermediate
values of $J\dom$ the curve crosses the DMRG result and predicts critical
couplings, which must be considered as too low. It is interesting that the
result of Ref.\@ \cite{uhrig98b} gives a good description of $\gc$ up to
$J\dom\approx 10$ although it is systematic only to order $g^2\dom^2$ and
$g^2J\dom^3$.
We presume that the two leading orders comprise already the main competition
between dimerization and delocalization. Further orders start to build in
resonance effects. But they do not succeed for all $J\dom$ because the
effective interactions are restricted to small distances only in our ansatz.
It is not uncommon (cf.\@ mean field treatments) that the leading orders yield
physically reasonable though not accurate results whereas systematic
improvements are difficult to implement (see Ref.\@ \cite{uhrig96b} for an
expansion in the inverse coordination number as example).

The inclusion of the next order in the systematic expansion by Wei\ss{}e et
al.\@ leads to slight improvements for $J\dom<0.15$ which is an estimate of
the convergence radius of the series expansion. But the higher order terms
produce too large critical couplings even for intermediate values of $J\dom$
\cite{weisse99bremark}. Furthermore, Wei\ss{}e et al.\@ used the Lang-Firsov
transformation to combine adiabatic and antiadiabatic effects in their theory.
This leads actually to an additional increase of the critical couplings for
intermediate $J\dom$ (cf.\@ Fig.~6 in Ref.\@ \onlinecite{weisse99b}). We
conclude that the extended flow equation approach used here has important
advantages over the systematic series expansion. This is due to its
renormalizing property, i.\,e.\@ the feedback of the $\ell$-dependence of all
couplings. In this way the region of validity is increased up to $J\dom\approx
0.35$.

For values of $J\dom<1$ Sun et al.\@ use $\gc\dom=\sqrt{J/(2\om)}$
\cite{sun00,sun00remark}. This is smaller by about a factor $1/\sqrt{2}$ than
the first order result obtained by Kuboki and Fukuyama \cite{kubok87}, who
find $\gc\dom=\sqrt{\alc/(1/2-\alc)\cdot J\dom}\approx\sqrt{J\dom}$. As can be
seen from Figure \ref{figgcvsjother}, the results of Sun et al.\@ do not
compare so well with those of the other approaches. For $J\dom>2$ they find
critical couplings $\gc\dom\approx 0.6$, which are about twice as large as the
$\gc$-values computed by DMRG \cite{bursi99}. For increasing $J\dom$, Sun et
al.\@ predict decreasing $\gc$ above $J/\omega\approx 5$ (not shown) in
contrast to the DMRG data.

While for not too large values of $J\dom$ the extension of the flow equation
approach to higher orders in $g\dom$ provides a fast method to retrieve data
as precise as DMRG calculations, the large $J\dom$ regime cannot be described
so easily. At $J\dom\approx 0.77$ the effective Hamiltonian is no longer
hermitian and the description breaks down. The next systematic extension is to
enlarge the operator subspace for $\opail$ to order $\opl^2$. The result once
again improves the calculations in order $\opl$ for small $J\dom$, but cannot
circumvent the ``downturn'' at $J\dom\approx 0.4$. We will argue below that
the inclusion of longer range interactions is necessary for $J\dom>0.35$.

\section{Modifications}
\label{capMod}
To enlarge the region of validity of the flow equation approach to larger
$J$s, we introduce some modifications. In a first attempt the restriction to a
given operator subspace can be dropped for intermediate terms arising in
(\ref{eqDGLA}-\ref{eqDGLJ12}) via $\opl\opai$. This circumvents the downturn
of the phase transition line but leads to too large critical couplings
$\gc(J)$. As these modifications changed the solutions of the differential
equations drastically the influence of longer range interactions present in
these intermediate terms can be judged as important. To include them in a
systematic way two additional couplings $J_3$ and $J_4$ are introduced.  They
correspond to spin interactions $\spr{}{+3}$ and $\spr{}{+4}$. The flow
equations for $J_3\ldep$ and $J_4\ldep$ can be extracted from
(\ref{eqDGLJ12}). As the critical coupling $\alc$ is only known for the
$J_1$-$J_2$-model we combine these four couplings to two effective couplings
$J_1^{\sw}\ldep$ and $J_2^{\sw}\ldep$ and thus
$\al^{\sw}\ldep=J_2^{\sw}/J_1^{\sw}$. This is done by using linear spin wave
theory \cite{auerb94}. The effective couplings $J_1^{\sw}$ and $J_2^{\sw}$ are
chosen such that the spin wave dispersion with $J_1^{\sw}$ and $J_2^{\sw}$
equals the one for $J_1$, $J_2$, $J_3$ and $J_4$. Thus we obtain
\begin{mathletters}
\begin{eqnarray}
  \label{eqJ1SW}
  J_1^{\sw} &=& \sqrt{(J_1+9J_3)(J_1+J_3)}\\
  \label{eqJ2SW}
  J_2^{\sw} &=& \sqrt{\frac{J_1+J_3}{J_1+9J_3}}(J_2+4J_4)\\
  \al^{\sw} &=& \frac{J_2^{\sw}}{J_1^{\sw}} = \frac{J_2+4J_4}{J_1+9J_3}\ .
  \label{eqAlphaSW}
\end{eqnarray}
\end{mathletters}
The curve of Fig.~\ref{figgcvsjour} denoted with $\al^{\sw}$ is the solution of
$\alc-\al^{\sw}_{\limellinf}=0$. The downturn does not occur any more,
but the critical couplings are too large. As $\diffell{J_3}$ and
$\diffell{J_4}$ are solved separately from all other coupling constants,
i.\,e.\@ $J_3$ and $J_4$ do not couple to the other differential equations,
another method denoted with $\diff^{\sw}$ is introduced. Differentiating
(\ref{eqJ1SW}) and (\ref{eqJ2SW}) one finds $\diffell{J_1^{\sw}}$ and
$\diffell{J_2^{\sw}}$. These two derivatives do depend on all $J_n$ and
$\diffell{J_n}$ ($n=1,\ldots,4$). They are calculated in each Runge-Kutta
step. This provides a coupling of $J_3$ and $J_4$ to the other flow equations.
The results from this procedure show again the downturn at $J\domz\approx 0.4$
and produce worse data for small $J\dom$ than the unmodified approach.

The inclusion of long range interaction terms modifies the results
drastically. So they are important and have to be treated correctly. It seems
that one has to extend the flow equation approach to very high orders in the
Liouville operator $\opl$, i.\,e.\@ the subspace for the coupling operator
$\opail$ should be fairly large, if one wants accurate data for large values
of $J\dom$. For $J\dom<0.35$, however, the method presented works very
accurately.

Due to the complicated spin commutators and products in the flow equations an
extension to higher orders in $\opl$ is cumbersome. A further hint of how this
could be achieved more easily shall be given. Recently non-standard choices of
the generator $\eta$ have been used successfully \cite{mielk98,knett00a}.
The main point is to introduce a quantum number $\qz$ and define the generator
via $\eta\ldep = [\qz,\hl]$.  Thus $[\he,\qz]=0$, i.\,e.\@ $\qz$ is conserved
with respect to $\he$. For our problem we introduce the phonon number
$\qz=\sum_i \bd_i b_i$. With this choice the generator reads
$\eta=\sum_i(\opai\bdi-\opadi\bi)$ and does not contain the Liouville
operator. The flow equations (for $T=0$) are pretty simple
\begin{mathletters}
\begin{eqnarray}
  \label{eqFlussA}
  \diffell{\opai} &=& -(\opl+\om)\opai\\
  \label{eqFlussW}
  \diffell{\om} &=& 2\langle[\opai,\opadi]\rangle\\
  \label{eqFlussJ}
  [\eta,\hsb] &=& -2\sum_i\opadi\opai\ .
\end{eqnarray}
\end{mathletters}
We investigated these flow equations in order $\opl$ and $\opl^2$. The results
are similar to those of the standard choice of $\eta$. Although the standard
choice (\ref{eqEtaGen}) leads to slightly better results for small couplings
$J\dom$, the alternative generator could be used to investigate very large
subspaces for $\opail$, as only $\opl$ enters linearly in (\ref{eqFlussA}) and
it is absent in (\ref{eqFlussW},\ref{eqFlussJ}).

\section{Summary}
\label{capSum}
We used the flow equation method to rotate the spin-phonon coupling of a
spin-Peierls Hamiltonian away. In this way we derived effective spin and
phonon models. The spin-phonon coupling induces longer range interactions,
i.\,e.\@ produces a non-vanishing frustration, and renormalizes the original
couplings. The effective phonon frequency is enlarged. This corresponds to the
phonon hardening observed for CuGeO$_3$.
The phase diagram describing the transition from a gapped to an ungapped state
is compared with DMRG results. For large values of $J\dom$ the results
deviate. In this regime very long range interactions appear to be
relevant and the transformation of the spin-Peierls chain onto a rather local
effective spin model must be considered as not sufficient.
The flow equation and the DMRG results coincide for $J\dom<0.35$.

\section{Acknowledgements}
The authors are indebted to R.~J.~Bursill for providing additional DMRG data
and acknowledge helpful discussions with H.~Fehske, U.~L\"ow,
E.~M\"uller-Hartmann and A.~Wei\ss{}e. This work was supported by the DFG
through SFB 341 and SP 1073.


\begin{thebibliography}{10}
\bibitem{uhrig98bremark}
For the $\gc(J)$-results of Ref.\@ \cite{uhrig98b} we use the version given in
  \cite{weisse99b}, which is corrected by a prefactor 1/2.
\bibitem{weisse99bremark}
For the fourth order perturbation theory curve in Fig.~\ref{figgcvsjother} we
  use formula (32-34) of Ref.\@ \cite{weisse99b}.
\bibitem{sun00remark}
To translate the notation of Ref.\@ \cite{sun00} into ours, one has to use
  $g\dom=(J\dom)^{3/2}\gamma/\sqrt{2}$. Thus, the straight line in Fig.~2 of
  Ref.\@ \cite{sun00} $\gamma=\om/J$ (for $J\dom<1$) corresponds to
  $g\dom=\sqrt{J/(2\om)}$.
\end{thebibliography}

\end{document}